\documentclass[12pt]{article}

\usepackage{amsmath}
\usepackage{cite}
\usepackage{amsfonts}
\usepackage{amssymb}
\def\eptwo{\left\{ \phantom{|}^{\mu\nu}_{ab} \right\}}
\def\epthree{\left\{ \phantom{|}^{\mu\nu\alpha}_{abc} \right\}}

\begin{document}

\title{Gauge invariant Lagrangian formulation \\
of massive higher spin fields in $(A)dS_3$ space}

\author{I.L. Buchbinder${}^{a}$\thanks{joseph@tspu.edu.ru},
T.V. Snegirev${}^a$\thanks{snegirev@tspu.edu.ru}, Yu.M.
Zinoviev$^b$\thanks{Yurii.Zinoviev@ihep.ru}
\\[0.5cm]
\it ${}^a$Department of Theoretical Physics,\\
\it Tomsk State Pedagogical University,\\
\it Tomsk 634061, Russia\\[0.3cm]
\it ${}^b$Institute for High Energy Physics,\\
\it Protvino, Moscow Region, 142280, Russia}
\date{}

\maketitle

\begin{abstract}
We develop the frame-like formulation of massive bosonic higher
spins fields in the case of 3-dimensional $(A)dS$ space  with the
arbitrary cosmological constant. The formulation is based on
gauge-invariant description by involving the  Stueckelberg auxiliary
fields. The explicit form of the Lagrangians and the gauge
transformation laws are found. The theory can be written in terms of
gauge-invariant objects  similar to the massless theories, thus
allowing us to hope to use the same methods for investigation of
interactions. In the massive spin 3 field example we are able to
rewrite the Lagrangian using the new the so-called separated
variables, so that the study of Lagrangian formulation reduces to
finding the Lagrangian containing only half of the fields. The same
construction takes places for arbitrary integer spin field as well.
Further working in terms of separated variables, we build Lagrangian
for arbitrary integer spin and write it in terms of gauge-invariant
objects. Also, we demonstrate how to restore the full set of
variables, thus receiving Lagrangian for the massive fields of
arbitrary spin in the terms of initial fields.
\end{abstract}

\thispagestyle{empty}
\newpage
\setcounter{page}{1}

\section{Introduction}

The construction of the consistent theory describing the interacting
dynamics of higher spin fields is an old and intriguing problem that
really very far to be solved (for some recent reviews see
\cite{reviews}). Taking into account complexity of the problem and
technical difficulties it seems that higher spin theories in three
dimensions that appear to be much simpler may provide nice
playground to gain some useful experience. In particular, contrary
to the situation in $d \ge 4$ dimensions \cite{Vasiliev1}, in three
dimensions it is not necessary to consider infinite number of
massless higher spin fields to build a consistent interacting theory
\cite{Deser,Blencowe,CFPT}. Also it is important that similar to
gravity and supergravity theories in three dimensions \cite{TW},
many higher spin models can be considered as Chern-Simons ones
\cite{Blencowe,CFPT} and such possibility of rewriting the action in
the Chern-Simons form is widely used in investigations of possible
higher spin interactions.

Till now most of works on the three dimensional higher spin theories
were devoted to construction of interaction for massless fields
\cite{Blencowe,Prokushkin,CFPT} or parity odd topologically massive
ones \cite{TMG} with the most notable exception being the so called
New massive gravity \cite{NMG} (see also \cite{Zin12}). But massless
higher spin fields in three dimensions being pure gauge do not have
any physical degrees of freedom, while topologically massive one
though do contain physical degrees of freedom but due to their
specific properties can hardly be generalized to higher dimensions.
That is why we think that it is important to investigate parity even
interacting theories for massive higher spin fields. On the one
hand, such theories will certainly have physical interest by
themselves and from the other hand we may expect that they admit
higher dimensional generalizations.

Lagrangian description for massive higher spin fields in three
dimensions for the first time was given in \cite{TV}. In this paper
we provide frame-like gauge invariant formalism for such fields,
which in our opinion is the most suitable one for investigation of
massive higher spin field interactions. Recall that frame-like
formalism, generalizing the well known frame formalism for gravity,
was initially formulated for massless fields \cite{Vasiliev} and
then extended to massive case \cite{Zinoviev,PoV}. Such formalism
nicely works both in flat Minkowski space as well as in $(A)dS$
spaces with arbitrary value of cosmological constant. Also it gives
a possibility to use for the massive fields the results obtained for
more simple case of massless fields. Last but not least, it is the
application of frame-like formalism that allows reformulating the
massless three dimensional theories in the Chern-Simons form
\cite{CFPT}. In this work we will pay special attention to the
possibility to write Lagrangians for massive fields in terms of
gauge invariant objects and the Chern-Simons actions.

The work is organized as follows. In the rest part of the
introduction we formulate our notations and conventions as well as
give basic information on the frame-like formulation of higher spin
fields. In section 2 we briefly review the massless bosonic fields
in the frame-like formulation confining ourselves by main results
and the possibility to reformulate these theories in terms of
separated variables only. In section 3 we consider in detail the
example of massive spin 3 field. At first, we give explicit
expressions for the Lagrangian and gauge transformations and then we
introduce a gauge-invariant objects in terms of which Lagrangian is
rewritten in the Chern-Simons form. Also we show that it is possible
to introduce separated variables in a way similar to the massless
case. In section 4 we consider massive fields with an arbitrary
integer spin. This time, based on experience obtained in spin 3
case, we, from the very beginning, work with separated variables and
the Lagrangian written in the Chern-Simons form so that construction
turns out to be straightforward and rather simple. And then we
easily obtain the complete theory in terms of initial fields.

{\bf Notations and conventions.} In the frame-like formalism the
fields of the arbitrary integer spin are described by a pair of
1-forms\footnote{In $d \ge 4$ dimensions one also has to introduce
so called extra fields $\Omega_\mu{}^{b_1 \dots b_t, a_1 \dots
a_{s-1}}$, $2 \le t \le s-1$. But in three dimensions all these
fields are absent and this greatly simplifies all calculations.}
$$
\Phi_\mu{}^{a_1a_2...a_{s-1}}, \qquad
\Omega_\mu{}^{b,a_1a_2...a_{s-1}},\qquad
\Omega_\mu{}^{(b,a_1a_2...a_{s-1})}=0
$$
Here Greek letters are used for the world indices and Latin letters
for the local ones. The fields are totally symmetric with respect to
indices $a$ and traceless in all local indices. Indices in
parentheses here and below will mean a complete symmetrization
without normalization factor. The first field generalizes the frame
field in gravity and plays the role of the physical field and the
second one is auxiliary generalizing the concept of the Lorentz
connection in gravity. In $d=3$ there is a fully antisymmetric third
rank tensor $\varepsilon^{abc}$, using which we can introduce the
dual field
$$
\Omega_\mu{}^{a_1a_2...a_{s-1}} = \varepsilon^{bc(a_1}
\Omega_\mu{}^{b,c|a_2...a_{s-1})}
$$
which is completely symmetric and traceless on local indices.

We work in the $(A)dS$ space and use the notation $D_\mu$ for
covariant derivative normalized as follows
$$
[D_\mu,D_\nu] \xi^a = \lambda^2 e_{[\mu}{}^a \xi_{\nu]},
\qquad \lambda^2 = - \Lambda
$$
where $e_\mu{}^a$ plays the role of (non-dynamical) frame of $(A)dS$
background, and $\Lambda$ is an arbitrary cosmological constant. One
can always consider the situation with a positive or negative
$\Lambda$, as well as analyze the flat limit, putting $\lambda=0$.
Using the frame $e_\mu{}^a$ and its inverse, whenever it is
convenient we convert local indices into world ones and back, for
example
$$
\Phi_{\mu,\nu}{}^{a_1...a_{s-2}} = e_\nu{}^{a_{s-1}}
\Phi_{\mu}{}^{a_1...a_{s-1}}
$$
We will write down the world indices explicitly, so to write the
expression in totally antisymmetric form on the world indices (which
is equivalent to an external product of 1-forms) we will often use
the notation
$$
\eptwo = e^\mu{}_a e^\nu{}_b - e^\mu{}_b e^\nu{}_a
$$
and similarly for $\epthree$.

\section{The kinematics of massless fields}

In this section we give frame-like formalism for the massless fields
in $(A)dS_3$ space which then will be used as building blocks in
gauge invariant description for massive ones. \\[0.5cm]
{\bf Spin 2} This particle is described by a pair of fields
$\Omega_\mu{}^a$, $f_\mu{}^a$ with the following free Lagrangian
\begin{equation}
{\cal{L}} = \frac12 \eptwo \Omega_\mu{}^a \Omega_\nu{}^b -
\varepsilon^{\mu\nu\alpha} \Omega_\mu{}^a D_\nu f_\alpha{}^a +
\frac{\lambda^2}{2} \eptwo f_\mu{}^a f_\nu{}^b
\end{equation}
which is invariant under the following gauge transformations
\begin{equation}
\delta \Omega_\mu{}^a = D_\mu \eta^a + \lambda^2
\varepsilon_\mu{}^{ab} \xi^b, \qquad
\delta f_\mu{}^a = D_\mu \xi^a + \varepsilon_\mu{}^{ab} \eta^b
\end{equation}
For all massless higher spin fields $(s\geq 2)$ the free Lagrangians
can be conveniently rewritten using gauge invariant objects. For
$s=2$ we introduce the dual curvature and torsion
\begin{eqnarray}
{\cal{F}}_{\mu\nu}{}^{a} &=& D_{[\mu} \Omega_{\nu]}{}^{a} +
\lambda^2 \varepsilon_{[\mu}{}^{ab} f_{\nu]}{}^b \nonumber \\
{\cal{T}}_{\mu\nu}{}^{a} &=& D_{[\mu} f_{\nu]}{}^{a} +
\varepsilon_{[\mu}{}^{ab} \Omega_{\nu]}{}^b
\end{eqnarray}
With the help of these object the Lagrangian can be rewritten as
\begin{equation}
{\cal{L}} = - \frac14 \varepsilon^{\mu\nu\alpha} [\Omega_\mu{}^a
{\cal{T}}_{\nu\alpha}{}^{a} + f_\mu{}^a {\cal{F}}_{\nu\alpha}{}^{a}]
\end{equation}
In $AdS_3$ space with non-zero cosmological constant we can
reformulate theory in terms of new separated variables
\begin{equation}
\Omega_{\pm\mu}{}^a = \Omega_\mu{}^a \pm \lambda f_\mu{}^a, \qquad
\eta_\pm{}^a = \eta^a \pm \lambda \xi^a
\end{equation}
and corresponding gauge invariant objects
\begin{equation}
{\cal{F}}_{\pm\mu\nu}{}^{a} = {\cal{F}}_{\mu\nu}{}^{a} \pm
\lambda {\cal{T}}_{\mu\nu}{}^{a} = D_{[\mu} \Omega_{\pm\nu]}{}^{a} \pm
\lambda \varepsilon_{[\mu}{}^{ab} \Omega_{\pm\nu]}{}^b
\end{equation}
Then the original Lagrangian in the new field variables has the form
\begin{equation}
{\cal{L}} = - \frac{1}{8\lambda} \varepsilon^{\mu\nu\alpha}[
\Omega_{+\mu}{}^a {\cal{F}}_{+\nu\alpha}{}^{a} - \Omega_{-\mu}{}^a
{\cal{F}}_{-\nu\alpha}{}^{a}]
\end{equation}
Thus it is clear that the variables are separated and each half of the
Lagrangian is invariant under its own gauge transformation of the same
type, namely
\begin{equation}
\delta \Omega_{\pm\mu}{}^a = D_\mu \eta_\pm{}^a \pm \lambda
\varepsilon_\mu{}^{ab} \eta_\pm{}^b
\end{equation}
Moreover, in such formulation the invariance of the Lagrangian can be
easily checked using differential identities for curvatures:
\begin{equation}
D_{[\mu} {\cal F}_{\pm \nu\alpha]}{}^a = \mp \lambda
\varepsilon_{[\mu}{}^{ab} {\cal F}_{\pm \nu\alpha]}{}^b
\end{equation}
Thus, we can work with one half of the Lagrangian and one field only.
\\
{\bf Arbitrary spin $s$.} Here we will use a pair of fields
$\Omega_\mu{}^{a_1a_2...a_{s-1}}$ and $\Phi_\mu{}^{a_1a_2...a_{s-1}}$
which are completely symmetric and traceless in local indices. To
simplify expressions we will use compact notations
$\Phi_\mu{}^{a_1 \dots a_k} = \Phi_\mu{}^{(k)}$ and similarly for all
other objects where $k$ denotes the number of local indices. In these
notations the free Lagrangian has the form
\begin{equation}
{\cal{L}} = (-1)^s\big[ -\varepsilon^{\mu\nu\alpha}
\Omega_\mu{}^{(s-1)} D_{\nu} {\Phi}_{\alpha}{}^{(s-1)}
+ \frac{s-1}{2} \eptwo [ \Omega_\mu{}^{a(s-2)}
{\Omega}_{\nu}{}^{b(s-2)} + \lambda^2 \Phi_\mu{}^{a(s-2)}
{\Phi}_{\nu}{}^{b(s-2)} ] \big]
\end{equation}
while gauge transformations look like
\begin{eqnarray}
\delta{\Omega}_\mu{}^{(s-1)} &=& D_\mu \eta^{(s-1)}-\lambda^2
\varepsilon_\mu{}^{b(1} \xi^{s-2)b} \nonumber\\
\delta{\Phi}_\mu{}^{(s-1)} &=& D_\mu \xi^{(s-1)} -
\varepsilon_\mu{}^{b(1} \eta^{s-2)b}
\end{eqnarray}
As in the spin 2 case, the Lagrangian can be rewritten in terms of
gauge invariant objects (which we will call curvatures)
\begin{eqnarray}
{\cal{G}}_{\mu\nu}{}^{(s-1)} &=& D_{[\mu} {\Omega}_{\nu]}{}^{(s-1)} -
\lambda^2 \varepsilon_{[\mu}{}^{b(1} {\Phi}_{\nu]}{}^{s-2)b} \nonumber
\\
{\cal{H}}_{\mu\nu}{}^{(s-1)} &=& D_{[\mu} {\Phi}_{\nu]}{}^{(s-1)} -
\varepsilon_{[\mu}{}^{b(1} {\Omega}_{\nu]}{}^{s-2)b}
\end{eqnarray}
as
\begin{equation}
{\cal{L}} = - \frac{(-1)^s}{4} \varepsilon^{\mu\nu\alpha}[
{\Omega}_{\mu}{}^{(s-1)} {\cal{H}}_{\nu\alpha}{}^{(s-1)} +
{\Phi}_{\mu}{}^{(s-1)} {\cal{G}}_{\nu\alpha}{}^{(s-1)}]
\end{equation}
By analogy with spin 2 case we introduce the new variables
\begin{equation}
{\Omega}_{\pm\mu}{}^{(s-1)} = {\Omega}_{\mu}{}^{(s-1)} \pm \lambda
{\Phi}_{\mu}{}^{(s-1)}
\end{equation}
in this, the Lagrangian can be rewritten as follows
\begin{equation}
{\cal{L}} = - \frac{(-1)^s}{8\lambda} \varepsilon^{\mu\nu\alpha}[
{\Omega}_{+\mu}{}^{(s-1)} {\cal{G}}_{+\nu\alpha}{}^{(s-1)} -
{\Omega}_{-\mu}{}^{(s-1)} {\cal{G}}_{-\nu\alpha}{}^{(s-1)}]
\end{equation}
where
\begin{equation}
{\cal{G}}_{\pm\mu\nu}{}^{(s-1)} = {\cal{G}}_{\pm\mu\nu}{}^{(s-1)} \pm
\lambda {\cal{H}}_{\pm\mu\nu}{}^{(s-1)} = D_{[\mu}
{\Omega}_{\pm\nu]}{}^{(s-1)} \mp \lambda \varepsilon_{[\mu}{}^{b(1}
{\Omega}_{\pm\nu]}{}^{s-2)b}
\end{equation}
Each half of the Lagrangian is invariant with respect to one type of
gauge transformations
\begin{equation}
\delta \Omega_{\pm\mu}{}^{(s-1)} = D_\mu \eta_\pm{}^{(s-1)} \mp
\lambda \varepsilon_\mu{}^{b(1} \eta_\pm{}^{s-2)b}, \qquad
\eta_\pm{}^{(s-1)} = \eta^{(s-1)} \pm \lambda \xi^{(s-1)}
\end{equation}
Similarly to spin 2 case, the invariance of the Lagrangian in such
formulation directly follows from differential identities for
curvatures:
\begin{equation}
D_{[\mu} {\cal G}_{\pm \nu\alpha]}{}^{(k)} = \pm \lambda
\varepsilon_{[\mu}{}^{a(1} {\cal G}_{\pm \nu\alpha]}{}^{k-1)a}
\end{equation}
Thus as in the spin 2 case one can work with only one field and one
half of the Lagrangian, which in the component form looks like:
\begin{equation}
{\cal{L}} = \frac{(-1)^s}{4\lambda} [\lambda (s-1) \eptwo
\Omega_{+\mu}{}^{a(s-2)} {\Omega}_{+\nu}{}^{b(s-2)} -
\varepsilon^{\mu\nu\alpha} \Omega_{+\mu}{}^{(s-1)} D_{\nu}
{\Omega}_{+\alpha}{}^{(s-1)}]
\end{equation}
In the massless case it may seem that using the separated variables
does not simplify calculations very much. However, in a massive case
due to the large number of fields involved such technics gives
significant simplification.

\section{Massive spin 3 field}

In this section we begin with the usual gauge invariant description
of massive spin 3 field adopted to $d=3$ dimensions and then we show
that it can be reformulated in terms of separated variables similar
to the massless case.

Massive spin 3 field for the frame-like gauge invariant description
\cite{Zinoviev} requires four pairs of fields
$(\Omega_\mu{}^{ab},\Phi_\mu{}^{ab})$, $(\Omega_\mu{}^a,f_\mu{}^a)$,
$(B^a, A_\mu)$ and $(\pi^a,\varphi)$, the last three of which play
the roles of Stueckelberg fields. The Lagrangian has the form
\begin{eqnarray}
{\cal{L}}_0 &=& - \eptwo \Omega_\mu{}^{ac} \Omega_\nu{}^{bc} +
\varepsilon^{\mu\nu\alpha} \Omega_\mu{}^{ab} D_\nu \Phi_\alpha{}^{ab}
+ \frac12 \eptwo \Omega_\mu{}^a \Omega_\nu{}^b -
\varepsilon^{\mu\nu\alpha} \Omega_\mu{}^a D_\nu f_\alpha{}^a +
\nonumber \\
&& + \frac12 B^a B^a - \varepsilon^{\mu\nu\alpha} B_\mu D_\nu
A_\alpha - \frac12 \pi^a \pi^a + \pi^\mu D_\mu \varphi + \nonumber\\
&& + \varepsilon^{\mu\nu\alpha}[ 3m \Omega_{\mu,\nu}{}^a f_\alpha{}^a
+ m \Phi_{\mu,\nu}{}^a \Omega_\alpha{}^a - 2\tilde{m} \Omega_{\mu,\nu}
A_\alpha + \tilde{m} f_{\mu,\nu} B_\alpha] + M \pi^\mu A_\mu +
\nonumber\\
&& + \eptwo [-\frac{M^2}{36} \Phi_\mu{}^{ac} \Phi_\nu{}^{bc} +
\frac{M^2}{8} f_\mu{}^a f_\nu{}^b] + M\tilde{m} e^\mu{}_a f_\mu{}^a
\varphi + 3\tilde{m}^2 \varphi^2
\end{eqnarray}
and is invariant under the following set of gauge transformations
\begin{eqnarray}
\delta_0 \Omega_\mu{}^{ab} &=& D_\mu \eta^{ab} +
\frac{a_2}{2}(e_\mu{}^{(a} \eta^{b)} - \frac23 g^{ab} \eta_\mu) -
\frac{M^2}{36} \varepsilon_\mu{}^{c(a} \xi^{b)c} \nonumber \\
\delta_0 \Phi_\mu{}^{ab} &=& D_\mu \xi^{ab} - \varepsilon_\mu{}^{c(a}
\eta^{b)c} + \frac{3m}{2}(e_\mu{}^{(a} \xi^{b)} - \frac23 g^{ab}
\xi_\mu) \nonumber \\
\delta_0 \Omega_\mu{}^a &=& D_\mu \eta^a + 3m \eta_\mu{}^a +
\frac{M^2}{4} \varepsilon_\mu{}^{ab} \xi^b \\
\delta_0 f_\mu{}^a &=& D_\mu \xi^a + \varepsilon_\mu{}^{ab} \eta^b + m
\xi_\mu{}^a + 2\tilde{m} e_\mu{}^a \xi \nonumber \\
\delta_0 B^a &=& - 2\tilde{m} \eta^a, \qquad
\delta_0 A_\mu = D_\mu \xi + \tilde{m} \xi_\mu \nonumber \\
\delta_0 \pi^a &=& M\tilde{m} \xi^a, \qquad
\delta_0 \varphi = - M \xi \nonumber
\end{eqnarray}
where
$$
\tilde{m}^2 = 8m^2 + 4\lambda^2,
\qquad M^2 = 18(3m^2 + 2\lambda^2)
$$

Contrary to the massless case this Lagrangian can not be rewritten
in terms of separated variables. Indeed, while description of spin 3
and spin 2 components contains the symmetric pairs
($\Omega_\mu{}^{ab}$, $\Phi_\mu{}^{ab}$) and ($\Omega_\mu{}^a$,
$f_\mu{}^a$), it is not the case for spin 1 and spin 0 ones. To make
it possible we partially fix the gauge $\varphi=0$, solve the
algebraic equation for the $\pi$ field expressing
$A_\mu=\frac{1}{M}\pi_\mu$ and change the normalization
$\pi_\mu\rightarrow M\pi_\mu$, because now the field $\pi^a$ will
play the role of physical one. As a result, we obtain the Lagrangian
\begin{eqnarray}\label{S3_1}
{\cal{L}}_0 &=& - \eptwo \Omega_\mu{}^{ac} \Omega_\nu{}^{bc} +
\varepsilon^{\mu\nu\alpha} \Omega_\mu{}^{ab} D_\nu \Phi_\alpha{}^{ab}
+\frac12 \eptwo \Omega_\mu{}^a \Omega_\nu{}^b -
\varepsilon^{\mu\nu\alpha} \Omega_\mu{}^a D_\nu f_\alpha{}^a +
\nonumber\\
&& + \frac12 B^a B^a - \varepsilon^{\mu\nu\alpha} B_\mu D_\nu
\pi_\alpha + \nonumber\\
&& + \varepsilon^{\mu\nu\alpha}[ 3m\Omega_{\mu,\nu}{}^a f_\alpha{}^a +
m \Phi_{\mu,\nu}{}^a \Omega_\alpha{}^a - 2\tilde{m} \Omega_{\mu,\nu}
\pi_\alpha + \tilde{m} f_{\mu,\nu} B_\alpha] + \nonumber\\
&& + \eptwo [-\frac{M^2}{36} \Phi_\mu{}^{ac} \Phi_\nu{}^{bc} +
\frac{M^2}{8} f_\mu{}^a f_\nu{}^b] + \frac{M^2}{2} \pi^a \pi^a
\end{eqnarray}
The gauge transformations take the form
\begin{eqnarray}
\delta_0 \Omega_\mu{}^{ab} &=& D_\mu \eta^{ab} +
\frac{m}{2}(e_\mu{}^{(a} \eta^{b)} - \frac23 g^{ab} \eta_\mu) -
\frac{M^2}{36} \varepsilon_\mu{}^{c(a} \xi^{b)c} \nonumber \\
\delta_0 \Phi_\mu{}^{ab} &=& D_\mu \xi^{ab} - \varepsilon_\mu{}^{c(a}
\eta^{b)c} + \frac{3m}{2} (e_\mu{}^{(a} \xi^{b)} - \frac23 g^{ab}
\xi_\mu) \nonumber \\
\delta_0 \Omega_\mu{}^a &=& D_\mu \eta^a + 3m \eta_\mu{}^a +
\frac{M^2}{4} \varepsilon_\mu{}^{ab} \xi^b \\
\delta_0 f_\mu{}^a &=& D_\mu \xi^a + \varepsilon_\mu{}^{ab} \eta^b + m
\xi_\mu{}^a \nonumber \\
\delta_0 B^a &=& - 2\tilde{m} \eta^a, \qquad
\delta_0 \pi^a = \tilde{m} \xi^a \nonumber
\end{eqnarray}
Having at our disposal the explicit form of gauge transformations
we can build six gauge invariant objects (which we will call
curvatures though there are two-forms as well as one-forms among them)
\begin{eqnarray}
{\cal{G}}_{\mu\nu}{}^{ab} &=& D_{[\mu} \Omega_{\nu]}{}^{ab} +
\frac{m}{2} (e_{[\mu}{}^{(a} \Omega_{\nu]}{}^{b)} +
\frac23 g^{ab} \Omega_{[\mu,\nu]}) - \frac{M^2}{36}
\varepsilon_{[\mu}{}^{c(a} \Phi_{\nu]}{}^{b)c} \nonumber \\
{\cal{H}}_{\mu\nu}{}^{ab} &=& D_{[\mu} \Phi_{\nu]}{}^{ab} -
\varepsilon_{[\mu}{}^{c(a} \Omega_{\nu]}{}^{b)c} +
\frac{3m}{2} (e_{[\mu}{}^{(a} f_{\nu]}{}^{b)} +
\frac23 g^{ab} f_{[\mu,\nu]}) \nonumber \\
{\cal{F}}_{\mu\nu}{}^{a} &=& D_{[\mu}\Omega_{\nu]}{}^{a} -
3m \Omega_{[\mu,\nu]}{}^a - \tilde{m} e_{[\mu}{}^{(a} B_{\nu]} +
\frac{M^2}{4} \varepsilon_{[\mu}{}^{ab} f_{\nu]}{}^b \nonumber \\
{\cal{T}}_{\mu\nu}{}^{a} &=& D_{[\mu} f_{\nu]}{}^{a} +
\varepsilon_{[\mu}{}^{ab} \Omega_{\nu]}{}^b -
m \Phi_{[\mu,\nu]}{}^a + 2\tilde{m} e_{[\mu}{}^{(a} \pi_{\nu]} \\
{\cal{B}}_{\mu}{}^{a} &=& D_{\mu} B^{a} + 2\tilde{m} \Omega_\mu{}^a -
\frac{M^2}{2} \varepsilon_\mu{}^{ab} \pi^b - W_\mu{}^a \nonumber \\
\Pi_{\mu}{}^{a} &=& D_{\mu} \pi^{a} - \frac12 \varepsilon_\mu{}^{ab}
B^b - \tilde{m} f_\mu{}^a - V_\mu{}^a \nonumber
\end{eqnarray}
where zero form auxiliary fields $W^{ab}, V^{ab}$ (that do not enter
the free Lagrangian) are symmetric and transformed as follows
$$
\delta W^{ab} = 6m\tilde{m} \eta^{ab}, \qquad
\delta V^{ab} = - m\tilde{m} \xi^{ab}
$$
Then it is possible to rewrite the Lagrangian (\ref{S3_1}) using
these curvatures in the following form
\begin{equation}
{\cal{L}}_0 = \frac14 \varepsilon^{\mu\nu\alpha} [\Omega_\mu{}^{ab}
{\cal{H}}_{\nu\alpha}{}^{ab} + \Phi_\mu{}^{ab}
{\cal{G}}_{\nu\alpha}{}^{ab} - \Omega_\mu{}^{a}
{\cal{T}}_{\nu\alpha}{}^{a} - f_\mu{}^{a} {\cal{F}}_{\nu\alpha}{}^{a}
- 2 B_\mu{} {\Pi}_{\nu,\alpha} - 2 \pi_\mu{} {\cal{B}}_{\nu,\alpha}]
\end{equation}
Such description contains three symmetric pairs ($\Omega_\mu{}^{ab}$,
$\Phi_\mu{}^{ab}$), ($\Omega_\mu{}^a$, $f_\mu{}^a$) and ($B^a$,
$\pi^a$). Thus we can introduce new variables
\begin{align*}
&
\hat{\Omega}_{\pm\mu}{}^{ab} = \Omega_{\mu}{}^{ab} \pm
\frac{M}{6} \Phi_\mu{}^{ab}
&& \hat{B}_{\pm}^a = B^a \mp M\pi^a \\
&
\hat{\Omega}_{\pm\mu}{}^{a} = \Omega_\mu{}^{a} \pm
\frac{M}{2} f_\mu{}^{a}
&& \hat{W}_{\pm\mu}{}^a = W_\mu{}^a \mp MV_\nu{}^a
\end{align*}
while the corresponding curvatures will have the form
\begin{eqnarray}
\hat{\cal{G}}_{\pm\mu\nu}{}^{ab} &=& D_{[\mu}
\hat{\Omega}_{\pm\nu]}{}^{ab} + \frac{m}{2}[e_{[\mu}{}^{(a}
\hat{\Omega}_{\pm\nu]}{}^{b)} + \frac23 g^{ab}
\hat{\Omega}_{\pm[\mu,\nu]}] \mp \frac{M}{6}\varepsilon_{[\mu}{}^{c(a}
\hat{\Omega}_{\pm\nu]}{}^{b)c} \nonumber \\
\hat{\cal{F}}_{\pm\mu\nu}{}^{ab} &=& D_{[\mu}
\hat{\Omega}_{\pm\nu]}{}^{a} - 3m \hat{\Omega}_{\pm[\mu,\nu]}{}^a -
\tilde{m} e_{[\mu}{}^{a} \hat{B}_{\pm\nu]} \pm \frac{M}{2}
\varepsilon_{[\mu}{}^{ab} \hat{\Omega}_{\pm\nu]}{}^b \label{S3_2} \\
\hat{\cal{B}}_{\pm\mu}{}^{a} &=& D_{\mu} \hat{B}_\pm^{a} + 2 \tilde{m}
\hat{\Omega}_{\pm\mu}{}^a \pm \frac{M}{2} \varepsilon_\mu{}^{ab}
\hat{B}_\pm^b - \hat{W}_{\pm\mu}{}^a \nonumber
\end{eqnarray}
In terms of the new fields and curvatures the free Lagrangian can be
rewritten as
\begin{eqnarray}\label{S3_3}
{\cal{L}}_0 &=& \frac{1}{4M} \varepsilon^{\mu\nu\alpha}[
3\hat\Omega_{+\mu}{}^{ab} \hat{\cal{G}}_{+\nu\alpha}{}^{ab} -
\hat\Omega_{+\mu}{}^{a} \hat{\cal{F}}_{+\nu\alpha}{}^{a} +
\hat{B}_{+\mu} \hat{\cal{B}}_{+\nu\alpha}] - \nonumber\\
 && - \frac{1}{4M} \varepsilon^{\mu\nu\alpha}[
3\hat\Omega_{-\mu}{}^{ab} \hat{\cal{G}}_{-\nu\alpha}{}^{ab} -
\hat\Omega_{-\mu}{}^{a} \hat{\cal{F}}_{-\nu\alpha}{}^{a} +
\hat{B}_{-\mu} \hat{\cal{B}}_{-\nu\alpha}]
\end{eqnarray}
We see that the variables are separated, i.e. now there is no mixing
between them. Each half is invariant with respect to one type gauge
transformations and it looks like the Chern-Simons action for the
massless fields. Moreover, as in the massless case the gauge
invariance of the Lagrangian can be most easily checked using
corresponding differential identities:
\begin{eqnarray}
D_{[\mu} \hat{\cal G}_{\pm \nu\alpha]}{}^{ab} &=& - \frac{m}{2} [
e_{[\mu}{}^{(a} \hat{\cal F}_{\pm \nu\alpha]}{}^{b)} + \frac{2}{3}
g^{ab} \hat{\cal F}_{\pm [\mu\nu,\alpha]} ] \pm \frac{M}{6}
\varepsilon_{[\mu}{}^{c(a} \hat{\cal G}_{\pm \nu\alpha]}{}^{b)c}
\nonumber \\
D_{[\mu} \hat{\cal F}_{\nu\alpha]}{}^a &=& - 3m
\hat{\cal G}_{\pm [\mu\nu,\alpha]}{}^a + \tilde{m} e_{[\mu}{}^a
\hat{\cal B}_{\pm \nu,\alpha]} \mp \frac{M}{2}
\varepsilon_{[\mu}{}^{ab} \hat{\cal F}_{\pm \nu\alpha]}{}^b \\
D_{[\mu} \hat{\cal B}_{\pm \nu]}{}^a &=& 2\tilde{m}
\hat{\cal F}_{\pm\mu\nu}{}^a \mp \frac{M}{2} \varepsilon_{[\mu}{}^{ab}
\hat{\cal B}_{\pm \nu]}{}^b - \hat{\cal W}_{\pm\mu\nu}{}^a \nonumber
\end{eqnarray}
Thus we can work with one half of the fields only, in this component
form for the corresponding Lagrangian (with a plus sign omitted)
looks like:
\begin{eqnarray}\label{S3_4}
{\cal{L}}_0 &=& \frac{1}{4M} \big\{\varepsilon^{\mu\nu\alpha}[
6\hat\Omega_\mu{}^{ab} D_{\nu} \hat{\Omega}_{\alpha}{}^{ab}
-2\hat\Omega_\mu{}^{a} D_{\nu} \hat{\Omega}_{\alpha}{}^{a} +
\hat{B}_\mu D_{\nu}\hat{B}_\alpha + \nonumber\\
 && \qquad\qquad + 12m\hat{\Omega}_{\mu,\nu}{}^a
\hat\Omega_\alpha{}^{a} + 4\tilde{m} \hat\Omega_{\mu,\nu}
\hat{B}_{\alpha}] - \nonumber\\
 && \qquad - 2M\eptwo \hat\Omega_\mu{}^{ac} \hat{\Omega}_{\nu}{}^{bc}
+ M\eptwo \hat\Omega_\mu{}^{a} \hat{\Omega}_{\nu}{}^b + M\hat{B}^2
\big\}
\end{eqnarray}
This Lagrangian is invariant under the following gauge
transformations
\begin{eqnarray}
\delta_0 \hat{\Omega}_\mu{}^{ab} &=& D_\mu \hat{\eta}^{ab} +
\frac{m}{2} [e_\mu{}^{(a} \hat{\eta}^{b)} - \frac23 g^{ab}
\hat{\eta}_\mu] - \frac{M}{6} \varepsilon_\mu{}^{c(a} \hat{\eta}^{b)c}
\nonumber \\
\delta_0 \hat{\Omega}_\mu{}^a &=& D_\mu \hat{\eta}^a + \frac{M}{2}
\varepsilon_\mu{}^{ab} \hat{\eta}^b + 3m\hat{\eta}_\mu{}^a, \qquad
\delta_0 \hat{B}^a = - 2\tilde{m} \hat{\eta}^a \label{S3_5}
\end{eqnarray}
where the gauge parameters are defined as follows
$$
\hat{\eta}^{ab} = \eta^{ab} + \frac{M}{6} \xi^{ab}, \qquad
\hat\eta^a = \eta^a + \frac{M}{2} \xi^a
$$
Thus we have obtained a frame-like gauge invariant description of
massive spin 3 fields in terms of separated variables. Actually we
have two identical copies of the same Lagrangian. Therefore it is
sufficient to study a structure only one of them and restore a total
Lagrangian at the very end. Now we are ready for generalization to
the arbitrary spin field case. We will see that the similar
situation is also valid in general case.

\section{Massive field with arbitrary integer spin}

In this section we construct frame-like gauge invariant formulation
for massive fields with the arbitrary integer spin. Having gained
some experience from the spin 3 case, from the very beginning we
will work in terms of separated variables. Thus we will look for the
Lagrangian in the form
\begin{equation} \label{S4_1}
{\cal L}_0 = \sum_{k=1}^{s-1} (-1)^k \frac{a_k}{2}
\varepsilon^{\mu\nu\alpha} \hat{\Omega}_\mu{}^{(k)}
\hat{\cal G}_{\nu\alpha}{}^{(k)} + \frac{a_0}{2}
\varepsilon^{\mu\nu\alpha} \hat{B}_\mu \hat{\cal B}_{\nu,\alpha}
\end{equation}
while for the curvatures we will use the following expressions
generalizing that of (\ref{S3_2}):
\begin{eqnarray}
\hat{\cal G}_{\mu\nu}{}^{(k)} &=& D_{[\mu} \hat{\Omega}_{\nu]}{}^{(k)}
- \gamma_k \hat\Omega_{[\mu,\nu]}{}^{(k)} + \beta_k
\varepsilon_{[\mu}{}^{b(1} \hat{\Omega}_{\nu]}{}^{k-1)b} + \nonumber
\\
 && + \alpha_k [e_{[\mu}{}^{(1} \hat{\Omega}_{\nu]}{}^{k-1)} +
\frac{2}{2k-1} g^{(2} \hat{\Omega}_{[\mu,\nu]}{}^{k-2)}] \label{S4_2}
\\
\hat{\cal{B}}_\mu{}^a &=& D_{\mu} \hat{B}^{a} - \gamma_0
\hat{\Omega}_\mu{}^a + \beta_1 \varepsilon_\mu{}^{ba} \hat{B}^b -
\hat{W}_\mu{}^a \nonumber
\end{eqnarray}
In this, corresponding gauge transformations will look like (compare
(\ref{S3_5})):
\begin{eqnarray}
\delta_0 \hat{\Omega}_\mu{}^{(k)} &=& D_\mu \hat{\eta}^{(k)} +
\alpha_k [e_\mu{}^{(1} \hat{\eta}^{k-1)} - \frac{2}{2k-1} g^{(2}
\hat{\eta}_\mu{}^{k-2)}] + \nonumber \\
 && + \beta_k \varepsilon_\mu{}^{b(1} \hat{\eta}^{k-1)b} + \gamma_k
\hat{\eta}_\mu{}^{(k)} \label{S4_3} \\
\delta\hat{B}^a &=& \gamma_0 \hat\eta^a  \nonumber
\end{eqnarray}
First of all, note that in order to the Lagrangian rewritten in
terms of initial fields $\hat{\Omega}_\mu{}^{(k)} =
\Omega_\mu{}^{(k)} - \beta_k \Phi_\mu{}^{(k)}$ and $\hat{B}^a = B^a
+ \kappa_0 \pi^a$ had the canonically normalized kinetic terms, we
have to put
\begin{equation}
a_k = - \frac{1}{4\beta_k}, \quad k \ge 1, \qquad
a_0 = - \frac{1}{4\kappa_0}, \qquad \kappa_0 = 2\beta_1 \label{S4_4}
\end{equation}
Then one can show by direct calculations that for curvatures
(\ref{S4_2}) be invariant under gauge transformations (\ref{S4_3})
the following two relations must hold
\begin{eqnarray}
&& (k+2) \beta_{k+1} = k \beta_k \label{S4_5} \\
&& \frac{2k+3}{2k+1} \gamma_k \alpha_{k+1} - \gamma_{k-1} \alpha_k +
\beta_k{}^2 - \lambda^2 = 0 \label{S4_6}
\end{eqnarray}
From the explicit expressions for curvatures it is not hard to
obtain the differential identities
\begin{eqnarray}
D_{[\mu} \hat{\cal G}_{\nu\alpha]}{}^{(k)} &=& - \beta_k
\varepsilon_{[\mu}{}^{a(1} \hat{\cal G}_{\nu\alpha]}{}^{k-1)a} -
\gamma_k \hat{\cal G}_{[\mu\nu,\alpha]}{}^{(k)} - \alpha_k [
e_{[\mu}{}^{(1} \hat{\cal G}_{\nu\alpha]}{}^{k-1)} + \frac{2}{2k-1}
g^{(2} \hat{\cal G}_{[\mu\nu,\alpha]}{}^{k-2)} ] \nonumber \\
D_{[\mu} \hat{\cal B}_{\nu]}{}^a &=& - \beta_1
\varepsilon_{[\mu}{}^{ba} \hat{\cal B}_{\nu]}{}^b - \gamma_0
\hat{\cal G}_{\mu\nu}{}^a
\end{eqnarray}
Using them one can easily check that the Lagrangian will be gauge
invariant provided one more relation holds
\begin{equation}
k a_k \alpha_k - a_{k-1} \gamma_{k-1} = 0 \label{S4_7}
\end{equation}
Now let us solve these relations to obtain expressions for all
unknown coefficients. Equation (\ref{S4_5}) provides recurrent
relation on $\beta_k$ that allows expressing all of them in terms of
$\beta_{s-1}$
\begin{equation}
\beta_k = \frac{s(s-1)}{k(k+1)} \beta_{s-1}
\end{equation}
Then from (\ref{S4_7}) we obtain
\begin{equation}
\gamma_{k-1} = \frac{k(k+1)}{(k-1)} \alpha_k, \quad k \ge 2, \qquad
\gamma_0 = 2\alpha_1
\end{equation}
Substituting all these in (\ref{S4_5}), (\ref{S4_6}) we get:
\begin{eqnarray}
&& \frac{(2k+3)(k+1)(k+2)}{k(2k+1)} \alpha_{k+1}{}^2 -
\frac{k(k+1)}{(k-1)} \alpha_k{}^2 + \beta_k{}^2 - \lambda^2 = 0, \quad
k \ge 2 \label{S4_8} \\
&& 10 \alpha_2{}^2 - 2 \alpha_1{}^2 + \beta_1{}^2 - \lambda^2 = 0
\label{S4_9}
\end{eqnarray}
Taking into account that $\alpha_s = 0$ equation (\ref{S4_8}) for
$k = s-1$ gives us
\begin{equation}
\beta_{s-1}{}^2 = \frac{s(s-1)}{(s-2)} \alpha_{s-1}{}^2 + \lambda^2
\end{equation}
while for other values of $k$ it can be considered as recurrent
relation on $\alpha_k$. This relation can be directly solved and we
obtain
\begin{equation}
\alpha_k = \frac{(k-1)(s-k)(s+k)}{k^3(k+1)(2k+1)} \left[
\hat\alpha_{s-1}{}^2 + (s-k-1)(s+k-1) \lambda^2 \right], \qquad k > 1
\end{equation}
where we have introduced
$$
\hat\alpha_{s-1}{}^2 = \frac{s(s-1)^3}{s-2} \alpha_{s-1}{}^2
$$
while the remaining unknown coefficient $\alpha_1$ is determined by
(\ref{S4_9}).

The resulting Lagrangian can be written in component terms as
follows (compare (\ref{S3_4})):
\begin{eqnarray*}
{\cal L} &=& \sum_{k=1}^{s-1} \frac{(-1)^{k+1}}{4\beta_k} [
\varepsilon^{\mu\nu\alpha} ( \hat{\Omega}_\mu{}^{(k)} D_\nu
\hat{\Omega}_\alpha{}^{(k)} + 2k\alpha_k
\hat{\Omega}_{\mu,\nu}{}^{(k-1)} \hat{\Omega}_\alpha{}^{(k-1)} ) +
k\beta_k \eptwo \hat{\Omega}_\mu{}^{a(k-1)}
\hat{\Omega}_\nu{}^{b(k-1)} ] - \\
 && - \frac{1}{8\beta_1} [ \varepsilon^{\mu\nu\alpha} \hat{B}_\mu
D_\nu \hat{B}_\alpha - 2\beta_1 \hat{B}^a \hat{B}^a ]
\end{eqnarray*}

Now we can easily obtain the formulation in terms of initial
variables $\hat{\Omega}_\mu{}^{(k)} = \Omega_\mu{}^{(k)} - \beta_k
\Phi_\mu{}^{(k)}$ and $\hat{B}^a = B^a + 2\beta_1 \pi^a$. As we have
already noted, the Lagrangian will have a canonical form:
\begin{equation}
{\cal{L}} = \sum_{k=0}^{s-1} (-1)^k \frac14
\varepsilon^{\mu\nu\alpha} \big[\Omega_{\mu}{}^{(k)}
{\cal{H}}_{\nu\alpha}{}^{(k)} +
\Phi_{\mu}{}^{(k)}{\cal{G}}_{\nu\alpha}{}^{(k)}\big] - \frac12
\varepsilon^{\mu\nu\alpha} [{B}_\mu {\Pi}_{\nu\alpha} + {\pi}_\mu
{\cal{B}}_{\nu\alpha}]
\end{equation}
where gauge invariant curvatures are defined as follows
\begin{eqnarray}
{\cal{G}}_{\mu\nu}{}^{(k)} &=& D_{[\mu} {\Omega}_{\nu]}{}^{(k)} -
\frac{(k+1)(k+2)\alpha_k}{k} \Omega_{[\mu,\nu]}{}^{(k)} - \beta_k{}^2
\varepsilon_{[\mu}{}^{a(1}{\Phi}_{\nu]}{}^{k-1)a} + \nonumber \\
&& + \alpha_k[e_{[\mu}{}^{(1}{\Omega}_{\nu]}{}^{k-1)} +
\frac{2}{2k-1}g^{(2}{\Omega}_{[\mu,\nu]}{}^{k-2)}] \nonumber \\
{\cal{H}}_{\mu\nu}{}^{(k)} &=& D_{[\mu} {\Phi}_{\nu]}{}^{(k)} -
(k+1) \alpha_{k+1} \Phi_{[\mu,\nu]}{}^{(k)} -
\varepsilon_{[\mu}{}^{a(1} {\Omega}_{\nu]}{}^{k-1)a} + \nonumber \\
&& + \frac{\gamma_{k-1}}{k} [e_{[\mu}{}^{(1}
{\Phi}_{\nu]}{}^{k-1)} + \frac{2}{2k-1} g^{(2}
{\Phi}_{[\mu,\nu]}{}^{k-2)})] \\
{\cal{B}}_\mu{}^a &=& D_{\mu} {B}^{a} - 2\alpha_1 {\Omega}_\mu{}^a +
2\beta_1{}^2 \varepsilon_\mu{}^{ba}\pi^b - {W}_\mu{}^a \nonumber \\
{\Pi}_\mu{}^a &=& D_{\mu} \pi^a +\alpha_1 f_\mu{}^a + \frac{1}{2}
\varepsilon_\mu{}^{ba} {B}^b - V_\mu{}^a \nonumber
\end{eqnarray}
while gauge transformations leaving the Lagrangian and curvatures
invariant look like
\begin{eqnarray}
\delta \Omega_\mu{}^{(k)} &=& D_\mu \eta^{(k)} + \alpha_k
[e_\mu{}^{(1} \eta^{k-1)} - \frac23 g^{(2} \eta_\mu{}^{k-2)}] +
\nonumber \\
&& - \beta_k{}^2 \varepsilon_\mu{}^{b(1} \xi^{k-1)b} +
\frac{(k+1)(k+2)}{k} \alpha_k \eta_\mu{}^{(k)} \nonumber \\
\delta \Phi_\mu{}^{(k)} &=& D_\mu \xi^{(k)} + \frac{\gamma_{k-1}}{k}
[e_\mu{}^{(1} \xi^{k-1)} - \frac23 g^{(2} \xi_\mu{}^{k-2)}] - \\
&& - \varepsilon_\mu{}^{a(1} \eta^{k-1)a} + (k+1) \alpha_k
\xi_\mu{}^{(k)} \nonumber \\
\delta{B}^a &=& 2\alpha_1 \eta^a, \qquad
\delta\pi^a = - \alpha_1 \xi^a \nonumber
\end{eqnarray}
Let us also give here a component form for the Lagrangian to be
compared with (\ref{S3_4})
\begin{eqnarray}
{\cal{L}} &=& \sum_{k=1}^{s-1} (-1)^{k+1} \big[ \frac{k}{2} \eptwo
\Omega_{\mu}{}^{a(k-1)} {\Omega}_{\alpha}{}^{b(k-1)} -
\varepsilon^{\mu\nu\alpha} {\Omega}_{\mu}{}^{(k)} D_{\nu}
\Phi_{\alpha}{}^{(k)} + \nonumber \\
&& \qquad +\frac{k}{2} \beta_k{}^2 \eptwo \Phi_{\mu}{}^{a(k-1)}
{\Phi}_{\alpha}{}^{b(k-1)} \big] - \nonumber \\
&& - \sum_{k=2}^{s-1}(-1)^{k+1} \varepsilon^{\mu\nu\alpha}[ 
\gamma_{k-1} \Omega_{\mu,\nu}{}^{(k-1)} \Phi_{\alpha}{}^{(k-1)}
+ \alpha_k k {\Omega}_{\mu}{}^{(k-1)} \Phi_{\nu,\alpha}{}^{(k-1)}]
+ \nonumber \\
&& \qquad + \frac12 B^a{B}^a - \varepsilon^{\mu\nu\alpha} B_\mu
D_{\nu} \pi_\alpha - \varepsilon^{\mu\nu\alpha} [\alpha_1 B_\mu
f_{\nu,\alpha} - 2\alpha_1{\Omega}_{\mu,\nu} \pi_\alpha] +
2\beta_1{}^2 \pi^a \pi^a
\end{eqnarray}
As in terms of separated variables, there is one arbitrary parameter
$\alpha_{s-1}$ which is fixed by the normalization of the mass. For
example, the canonical normalization corresponds to
$$
\alpha_{s-1}{}^2 = \frac{s-2}{s(s-1)^2} m^2
$$

\section*{Summary}

In this paper we have given the gauge invariant Lagrangian
formulation of $d=3$ massive bosonic higher spin fields in $(A)dS_3$
space, working within the frame-like approach. The gauge invariance
is ensured by the use of the Stueckelberg auxiliary fields.
Similarly to the massless case we have rewritten the Lagrangians for
massive fields in terms of gauge invariant objects similarly to
Chern-Simons actions. This is achieved by a partial gauge fixing, so
that the scalar field is excluded. This was clearly demonstrated for
massive spin 3 field example with a further reformulation of the
theory in terms of separated field variables. The separated
variables allow us working with one half of fields that greatly
simplifies all calculations. Finally, using the separated variables,
we have constructed the Lagrangian for the massive field of
arbitrary integer spin and have shown how to restore the theory in
terms of original fields. Due to the partial gauge fixing, the
resulting theory in original fields does not contain the auxiliary
scalar field.

\section*{Acknowledgments}
I.L.B and T.V.S are grateful to the grant for LRSS, project No
224.2012.2 and RFBR grant, project No 12-02-00121-a for partial
support. Work of I.L.B was also partially supported by RFBR-Ukraine
grant, project No 11-02-9045. Work of Yu.M.Z was supported in parts by
RFBR grant No. 11-02-00814.

\end{document}